\documentclass[prl,reprint,groupedaddress]{revtex4-2}
\pdfoutput=1

 \usepackage{amsmath}
 \usepackage{amsfonts,amssymb,marginnote}
 \usepackage{graphicx}
 \usepackage{newtxtext,newtxmath}
\usepackage{dcolumn}

 \usepackage{hyperref}
 \hypersetup{
    colorlinks=true,       % false: boxed links; true: colored links
    linkcolor=blue,          % color of internal links
    citecolor=blue,        % color of links to bibliography
    filecolor=magenta,      % color of file links
    urlcolor=blue           % color of external links
}

\renewcommand{\epsilon}{\varepsilon}

\allowdisplaybreaks

\newcolumntype{d}[1]{D{.}{.}{#1}}

\let\originalleft\left
\let\originalright\right
\renewcommand{\left}{\mathopen{}\mathclose\bgroup\originalleft}
\renewcommand{\right}{\aftergroup\egroup\originalright}

\usepackage{comment}

\begin{document}

\frenchspacing

\title{Positron cooling via inelastic collisions in CF$_4$ and N$_2$ gases}
\author{A.~R. Swann}
\email{a.swann@qub.ac.uk}
\author{D.~G. Green}
\email{d.green@qub.ac.uk}
\affiliation{
%Centre for Theoretical Atomic, Molecular and Optical Physics,
School of Mathematics and Physics, Queen's University Belfast, University Road, Belfast BT7 1NN, United Kingdom}
\date{\today}

\begin{abstract}
Positron cooling via inelastic collisions in CF$_4$ and N$_2$ gases is simulated, including positron-positron interactions. 
Owing to the  molecular symmetries, cooling is assumed to be 
chiefly due to energy loss via vibrational (rotational) excitations for CF$_4$ (N$_2$). % gas molecules.  
For CF$_4$,  it is found that the inclusion of the dipole-inactive $\nu_1$  mode, in addition to the dipole-active modes $\nu_3$ and $\nu_4$, can provide room-temperature thermalization and an accurate cooling timescale. Combination cooling enabled by the $\nu_1$ mode, and positron-positron interactions both contribute to the Maxwellianization of the positron momentum distribution. 
%room-temperature thermalization
%including  including only the dipole-active modes $\nu_3$ and $\nu_4$ is insufficient, and that the $\nu_1$ dipole-inactive mode is crucial to achieve room-temperature thermalization and obtain an accurate cooling timescale.}
%\blue{positron-positron \dots}
For both gases the evolution of the positron temperature is found to be in excellent agreement with experiment.
\end{abstract}

\maketitle

The development of the  positron buffer-gas  trap in the 1980s \cite{Surko89} has enabled the routine trapping, accumulation, and delivery of positrons in beams \cite{Gilbert97,RevModPhys.87.247}; the study of low-energy antimatter-matter interactions with atoms and molecules, including scattering, binding, and annihilation \cite{0953-4075-38-6-R01,Gribakin10}; and the formation and exploitation of positronium \cite{Brawley:2010,cassidy2018} and antihydrogen \cite{alpha2}.
%
%\red{New first sentence to replace: Pioneering technological developments over the past few decades have resulted in the routine trapping, accumulation, and delivery of positrons in beams \cite{RevModPhys.87.247}, enabling the study of low-energy antimatter-matter interactions with atoms and molecules, including scattering, binding, and annihilation \cite{0953-4075-38-6-R01,Gribakin10}, and antihydrogen formation \cite{alpha2}. Most important} was the development of the modified Penning-Malmberg ``Surko'' trap \cite{Surko89}, first used to produce a positron beam in 1997 \cite{Gilbert97}, and now used in nearly all low-energy positron experiments worldwide. 
%
%In the typical setup, 
Positrons from a $^{22}$Na source (with energies 0--500~keV) are slowed to eV energies by passing through an $\sim$8~K solid-neon moderator, then magnetically guided into the three-stage buffer-gas trap. The first two stages %of the trap 
contain N$_2$ gas in which the positrons cool, typically through electronic excitation
 \footnote{
 N$_2$ is chosen as the buffer gas because it is the only known molecule for which the threshold for electronic excitation ($\approx$8.5~eV) is sufficiently below the threshold for positronium formation ($\approx$8.8~eV) \cite{Marjanovic16}.
 }. 
In the third stage, a mixture of N$_2$ and CF$_4$ (or SF$_6$)  is used to complete the thermalization of the positrons to room temperature, typically via rotational and vibrational excitation of the molecules  \cite{Marjanovic16}.
They can be cooled further in a cryogenic beam-tailoring trap which produces an ultra-high resolution ($\sim$7 meV FWHM) energy-tunable beam \cite{Natisin:2016}.
%After passing through an additional cryogenic stage, the positrons can be ejected as an ultra-high resolution ($\sim$7~meV FWHM) energy-tunable beam \cite{Natisin:2016}. 

Optimization of current traps, and development of next-generation traps, accumulators, beams and positron-based technologies require theoretical insight.
The theory of low-energy positron cooling in atomic and molecular gases, however, lags well behind experiment (see, e.g., Refs.~\cite{Griffith:1979,Charlton85} for early reviews). 
Most existing theoretical work has been for noble gases, for which the dominant positron-energy-loss mechanism is momentum transfer in elastic %positron-atom
  collisions. 
Solutions of the Fokker-Planck equation using model scattering cross sections \cite{Orth69,Campeanu77,Campeanu81,Campeanu82,Shizgal87,Boyle14} yielded limited agreement with experiment. 
Recently, however, a Monte Carlo approach employed by one of us that used accurate many-body-theory scattering cross sections %calculated using many-body-theory 
gave a complete description of positron cooling in noble gases, finding excellent agreement with experiment for cooling rates, time-dependent annihilation rates \cite{Green17}, and $\gamma$ spectra \cite{DGG_gamcool}.
%Positron 
Cooling at higher energies in noble gases via electronic excitation, ionization, and positronium formation has also been investigated \cite{Girardi-Schappo13}.
For molecular gases, less progress has been made, with even  cooling in N$_2$ and CF$_4$ not %quantitatively 
well understood, though some simulations exist \cite{Bankovic14,CF4_existing_note}.

Natisin \textit{et al.} \cite{Natisin14} performed measurements of positron cooling following microwave heating to $\sim$1500~K in CF$_4$, N$_2$, and CO gases, finding that 
their results were consistent with a positron momentum distribution (PMD) that remained Maxwellian throughout the cooling process. 
%For CF$_4$ this was perplexing, as 
For CF$_4$, their theoretical model (a simple %first-order 
differential equation for the mean energy loss, assuming a Maxwellian PMD throughout) included only the dominant $\nu_3$ mode: with this single vibrational channel, one should, however, expect the PMD to deviate significantly from Maxwellian as positrons below the vibrational excitation threshold cannot cool further, leading to a ``pileup'' of positrons just below the threshold.  The mechanism(s) causing ``Maxwellianization'' is not yet understood. % Thus, other mechanisms must be at play. 

Here, we calculate the evolution of the PMD and temperature during cooling in CF$_4$ and N$_2$ gases via inelastic collisions. % including positron-positron interactions. %via a Monte Carlo approach, simulating the experiments of Natisin \emph{et al.} \cite{Natisin14}. 
For CF$_4$, we show that even when both  dipole-active vibrational modes ($\nu_3$ and $\nu_4$)  are included \footnote{$\nu_3$ dominates due to its significantly larger cross section.}, pileups indeed occur, resulting in a non-Maxwellian PMD, and moreover, that the positrons do not even thermalize to room temperature. 
%Thus, another mechanism must be at play \red{[We need to at least provide a positron/gas number density at which positron-positron interactions should lead to improved `Maxwellization'. But, maybe we need to do the study via the code ourselves. At least this would remove any doubt. So we need to be clear on what the puzzle actually is.]}
We explore two  mechanisms that could effect Maxwellianization:

(1) Excitation of the dipole-inactive $\nu_1$ mode. We find that this has a significant effect on the cooling, providing a pathway for positrons below the lowest excitation threshold to continue to cool, mitigating the pileup and  leading to thermalization on the timescale observed in the experiment. It does not appear to be sufficient, however, to fully Maxwellianize the PMD.

(2) Positron-positron collisions.  The relative importance of positron-positron and positron-gas  collisions is governed by the ratio $R\equiv n_e/n_g$ of the positron number density $n_e$ to the gas  number density $n_g$. 
%, which governs %The ratio $R$ governs the relative importance of positron-positron and positron-gas  collisions. 
%
%The ratio $R\equiv n_e/n_g$, where $n_e$ and $n_g$ are the positron and gas number densities, respectively, 
While positron-positron collisions are known to be capable of effecting rapid Maxwellianization at high positron densities \cite{Trunec03}, it is not clear \emph{a priori} what magnitude of $R$ is required for this to occur. 
We find that the density ratio 
in the CF$_4$ cooling experiment of Natisin {\it et al.} \cite{Natisin14}, 
$R\sim10^{-7}$--$10^{-6}$ \cite{Surko_private}, may be %just above 
%at the level where positron-positron interactions can 
sufficient to noticeably enhance the Maxwellianization  of the PMD, beyond the effect of including the $\nu_1$ vibrational mode.
%As well as illuminating the role of the $\nu_1$ mode in CF$_4$, 

For N$_2$, we find that  the PMD remains Maxwellian during cooling via rotational excitations of the molecules, even without positron-positron collisions.

Overall, we obtain excellent agreement with experiment for both CF$_4$ and N$_2$. %, in turn providing a verification of the cross sections.
%\red{We've said a lot about CF$_4$ and hardly anything about N$_2$.}

\emph{Simulation procedure}.---%
The PMD $f(k,\tau)$, where $k$ is the  momentum and $\tau\equiv n_g t$ is the \textit{time-density} ($t$ being the time), normalized as $\int_0^\infty f(k,\tau)\,dk=1$, is calculated as follows. We use the {\tt ANTICOOL} program \cite{anticool}, modified to include vibrational and rotational inelastic positron-gas collisions, and positron-positron collisions. % via the method described in \cite{}. 
We employ a grid in $\tau$ with constant step size $\Delta \tau$. 
The initial momentum %\red{vector} 
of each positron is sampled from a Maxwell-Boltzmann distribution (MB) at $\sim$1500~K, corresponding to the experiment \cite{Natisin14}.
%\red{The probability of a positron-molecule collision is calculated as $P=W \, \Delta\tau$, where }
%\blue{
In each time-density step, and for each positron, the probability of a collision with either a molecule or another positron  is $P=W \, \Delta\tau$, where $W$ is the total  collision rate:
%When included, the total rate of  positrons with velocity $\mathbf v$ colliding with either a gas molecule or another positron is 
%\begin{equation}\label{eq:W}
$
W =  n_g [\int u {\sigma}_{eg}(u)f_g(\mathbf v')\,d\mathbf v'  + R\int u \sigma_{ee} (u )f_e(\mathbf v')\,d\mathbf v' ],
$
%\end{equation}
 where  $u\equiv\lvert \mathbf v - \mathbf v' \rvert $ is the relative speed of the incident positron and target gas molecule or positron, $\sigma_{eg}$ and $\sigma_{ ee}$  are the   positron-gas and positron-positron scattering cross sections, respectively,  $f_g(\mathbf v')$ and $f_e(\mathbf v')$ are the velocity distributions of the gas molecules and positrons, respectively,
%$W =  n_g \left(v_{eg} \bar{\sigma}_{\rm inel}(v_{eg}) + R v_{ee}\sigma_{ee}(v_{ee})\right)$ 
and $\Delta\tau$ is chosen such that $W \,\Delta\tau \ll1$ \footnote{In practice, we choose the time-density step $\Delta\tau$  small enough  that $W \,\Delta\tau <0.1$ is always satisfied.}. 
%To calculate $W$, %\red{in the positron-molecule term} 
We approximate  $\int u\sigma_{eg}(u)f_g(\mathbf v')\,d\mathbf v'$ by $v_{eg}\sigma_{eg}(v_{eg})$, where $v_{eg}$ is the relative speed of the positron and a single  gas molecule whose velocity is sampled from a MB at room temperature, $T_R=300$~K. To calculate %the positron-positron term in $W$, 
$\int u \sigma_{ee}(u)f_e(\mathbf v')\,d\mathbf v'$,
we use the  method of Weng and Kushner \cite{Weng90,supp_info}.
%$W=n_g v_\text{rel} \overline\sigma_\text{tot}(k_\text{rel})$
% is the collision rate, 
%Here $v_{eg}$ and $v_{ee}$ are the relative speeds of the incident positron and target gas molecule, and incident and target positron, respectively (see below), 
%For each $\tau$, and for each positron,  the velocity of the target gas molecule is sampled from a MB at room temperature,  $T_{\rm R}=300$~K. 
% $k_\text{eg}\equiv\mu v_\text{eg}$, $\mu$ is the reduced mass of the positron-molecule system and  $k_\text{ee}\equiv v_\text{ee}/2$
 % $v_\text{rel}$ is the relative speed of the two objects, 
%The simulation is implemented as follows. 
A random number $r_1\in[0,1]$ is drawn, and if $r_1<P$, then a collision is deemed to occur. 
%and the positron momentum updated as follows. %(otherwise, the positron momentum is unchanged). 
%If a collision occurs, 
If so, another random number 
$r_2\in[0,1]$ determines the target type:  if $r_2 < n_g v_{eg} {\sigma}_{eg}(v_{eg}) /W$, the target is a molecule; otherwise, it is  another positron. %\red{(If positron-positron collisions are not considered, i.e., $R=0$, then this step is skipped.)}
Finally, a random number $r_3\in[0,1]$  determines the specific scattering channel, i.e., the initial and final states of the molecule for an inelastic positron-gas collision, or the velocity of the target positron for  a positron-positron collision \cite{supp_info}.
%
%In the case of a positron-molecule collision, it remains to determine the initial and final states of the molecule (whose energy difference determines the inelastic energy change of the positron). 
%The total inelastic cross section is $\overline\sigma_\text{inel} =  \sum_{i,f} \overline\sigma_{i\to f}$, where $\overline\sigma_{i\to f}=p_i \sigma_{i\to f}$, with
%$\sigma_{i\to f}$ the positron-impact cross section  for the $i\to f$ (open-channel) molecular transition, and $p_i$ the Boltzmann-averaged occupation number  of the molecules in state $i$ in equilibrium.
%%If the collision is with a molecule 
%%The initial and final states $i$ and $f$ of the molecule are determined as follows.
% For clarity, let us combine the indices $i$ and $f$  into a single index $a$ that runs from 1 up to the product of the numbers of initial and final states of the molecule considered. Let
%$Q_a =\sum_{a'=1}^a \overline\sigma_{a'} (v_{eg})/\overline\sigma_\text{tot}(v_{eg})$.
%Another random number $r_3\in[0,1]$ is drawn, and the value of $a$ for which $Q_{a-1} \leq r_3 < Q_{a}$ gives the initial and final states of the molecule. 
%
%
In a positron-gas collision, the energy lost by the positron is $\varepsilon_f-\varepsilon_i$, where $\varepsilon_i$ and $\varepsilon_f$ are the energies of the initial and final states of the molecule, respectively;
% \red{ (if $\varepsilon_f<\varepsilon_i$ then the positron gains energy) [DO WE REALLY NEED TO SAY THIS?]};
in a positron-positron collision, the energy change of the incident positron is determined by the relative speed of the two positrons
%energy of the target positron %\red{[MORE PRECISELY, IS IT NOT RATHER THE DIFFERENCE IN ENERGIES OF THE INCIDENT AND TARGET POSITRONS?]} 
and the scattering angle \cite{Weng90}.
%\red{The positron's momentum vector is updated according to the energy change} \cite{supp_info}.
%The positron momentum is then updated by transforming to the positron-molecule center-of-mass frame, selecting the scattering plane and scattering angle  randomly \footnote{For simplicity, we assume the scattering to be isotropic.},  rotating the positron velocity appropriately and scaling its magnitude by the coefficient of restitution 
%%$\blue{e = } 
%$({1 - {2\,\Delta\varepsilon_{i\to f}}/{\mu v_\text{rel}^2}})^{1/2}$ 
%(where $\Delta\varepsilon_{i\to f}\equiv\varepsilon_f-\varepsilon_i$ is the energy transferred inelastically from the positron to the molecule,  $\varepsilon_i$ and $\varepsilon_f$ being the energies of molecular states $i$ and $f$, respectively), and transforming back to the lab frame.
%
%\blue{For positron-positron collisions we implement the approach of Wang and Kushner \cite{} (using their $\sigma_{ee}$ and energies) \dots}
%and $\sigma_{\rm ee}$ is the positron-positron scattering cross section, which we take to be that used in \cite{Kushner} \red{better ref/way of stating it?}.
%
The temperature of the positrons at a given $\tau$ is calculated as $T= k_{\rm rms}^2/3m_ek_B$, where $k_{\rm rms}$ is the root-mean-squared momentum of the positrons, $m_e$ is the positron mass, and $k_B$ is the Boltzmann constant.

\emph{Positron cooling in CF$_4$ gas}.---Since CF$_4$ has $T_d$ symmetry, %a spherical-top molecule,  
its electric dipole and quadrupole moments are zero. 
Thus, positron cooling in CF$_4$  is expected to be  predominantly via vibrational, rather than rotational, excitations of the molecules.  CF$_4$  possesses four fundamental vibrational modes. Modes $\nu_1$ and $\nu_2$, with energies $\varepsilon_1=113$~meV and $\varepsilon_2=53.9$~meV, are dipole inactive, while modes $\nu_3$ and $\nu_4$, with energies $\varepsilon_3=159$~meV and $\varepsilon_4=78.4$~meV, are dipole active \cite{CRC}. 
We investigate cooling of the positrons from 1700~K ($k_BT=146$~meV) to 300~K ($k_BT=26$~meV).
For simplicity, we  consider only fundamental transitions, and initially, we neglect the role of positron-positron collisions. % between the vibrational  ground state and a fundamental mode. 
We use the Born-dipole approximation  for the transition cross sections $\sigma_{0\rightleftarrows3,4}$ between the vibrational ground state $\nu_0$ and dipole-active modes $\nu_3$ and $\nu_4$ \cite{Itikawa74,supp_info}.
%\begin{eqnarray}
%\sigma_{0\to\nu}(k) =  \frac{8\pi g_\nu \mu_\nu^2}{k^2} \ln \frac{k + k_\nu^{(-)}}{k - k_\nu^{(-)}}, \label{eq:born_dip_ex} \\
%\sigma_{\nu\to0}(k) =   \frac{8\pi g_0 \mu_\nu^2}{k^2} \ln \frac{k_\nu^{(+)} + k}{k_\nu^{(+)} - k}  , \label{eq:born_dip_deex}
%\end{eqnarray}
%where $g_\nu$ is the degeneracy of mode $\nu$, $\mu_\nu$ is the transition dipole moment, and
%$k_\nu^{(\pm)} = (k^2 \pm 2\varepsilon_\nu)^{1/2}$, with $\varepsilon_\nu$  the energy of mode $\nu$ \cite{Itikawa74}.
%Modes $\nu_3$ and $\nu_4$  are  triply degenerate, %(i.e., $g_3=g_4=3$), 
%and we take $\varepsilon_3=5.84\times10^{-3}$~a.u. and $\varepsilon_4=2.88\times10^{-3}$~a.u. for the mode energies, and $\mu_3=0.12$~a.u. and $\mu_4=0.02$~a.u. for the transition dipole moments \cite{Bishop82}. The  ground state is nondegenerate. 
Semiempirical coupled-channel calculations  
\cite{Franz09} found $\sigma_{0\to3}$ to be similar in shape and magnitude to the corresponding Born-dipole calculation, but  $\sigma_{0\to4}$ to be approximately four times larger than the corresponding Born-dipole calculation (see Fig.~6 in Ref.~\cite{Franz09}).
Therefore, %we  use the Born-dipole cross sections, but for the $0\rightleftarrows4$ transitions 
we  also perform calculations using the Born-dipole  $\sigma_{0\to4}$ scaled by 4. 
For %the dipole-inactive mode 
$\nu_1$, we are unaware of any theoretical or experimental investigation of positron-impact excitation, so we turn to a set of %experimental 
measurements of $\sigma_{0\to1,3,4}$ for \textit{electron} impact \cite{Kurihara00} and  assume  the positron-impact  cross sections to be similar in shape and magnitude to the electron-impact ones (as predicted for  $\nu_3$ and $\nu_4$  \cite{Franz09}).
The overall shape of each measured cross section  is similar, differing in the threshold energy and  magnitude.
The peak in the measured $\sigma_{0\to1}$  is approximately 3.1 times higher than the peak in the  measured $\sigma_{0\to4}$ (see Fig.~1 in Ref.~\cite{Kurihara00}). %Thus, in the present work, even though the Born-dipole approximation cannot be invoked for a dipole-inactive mode, 
Thus, we estimate $\sigma_{0\rightleftarrows1}$ using the Born-dipole expressions for $\sigma_{0\rightleftarrows4}$, scaled by %an empirical factor of 
3.1. %Eqs.~(\ref{eq:born_dip_ex}) and (\ref{eq:born_dip_deex}) scaled by an empirical factor of 3.1, with $g_1\equiv g_4$, $\mu_1 \equiv \mu_4$, and $\varepsilon_1=4.15\times10^{-3}$~a.u.
We %also
 carry out a second set of calculations where we scale $\sigma_{0\rightleftarrows1}$ by a further factor of 4, to account for the fact that the Born-dipole calculation of $\sigma_{0\to4}$ is approximately four times smaller than  the corresponding coupled-channel calculation \cite{Franz09} (see above).
We are unaware of any existing calculations or measurements of $\sigma_{0\rightleftarrows2}$, so we neglect these transitions entirely.

\begin{figure}
\centering
\includegraphics*[width=0.89\columnwidth]{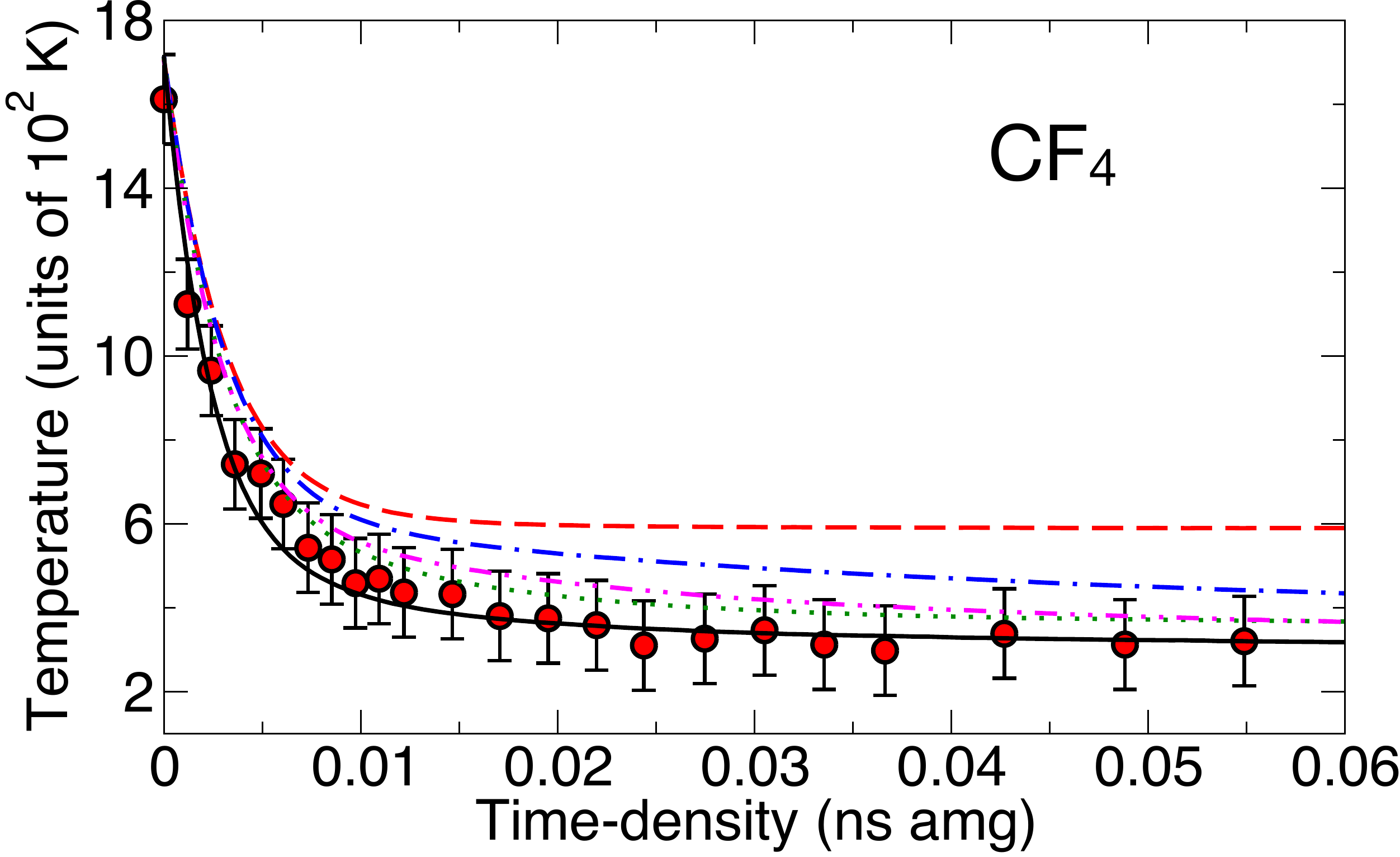}
\caption{\label{fig:CF4_temperature}
%Positron temperature   during cooling in CF$_4$. 
%Lines: dashed red, $\nu_3$; dot-dashed blue, $\nu_3+\nu_4$; dotted green, $\nu_3+4\nu_4$; dot-dash-dotted magenta, $\nu_3+\nu_4+\nu_1$; solid black, $\nu_3+4\nu_4+4\nu_1$  approximations (see text). Red circles: experimental data \cite{Natisin14}.
Positron temperature  during cooling in CF$_4$. 
Calculations in $\nu_3$ (dashed red),  $\nu_3+\nu_4$ (dot-dashed blue), $\nu_3+4\nu_4$ (dotted green),  $\nu_3+\nu_4+\nu_1$ (dot-dash-dotted magenta) and $\nu_3+4\nu_4+4\nu_1$ (solid black) approximations (see text), and experiment  \cite{Natisin14} (red circles).
}
\end{figure}

\begin{figure}
\centering
\includegraphics[width=0.875\columnwidth]{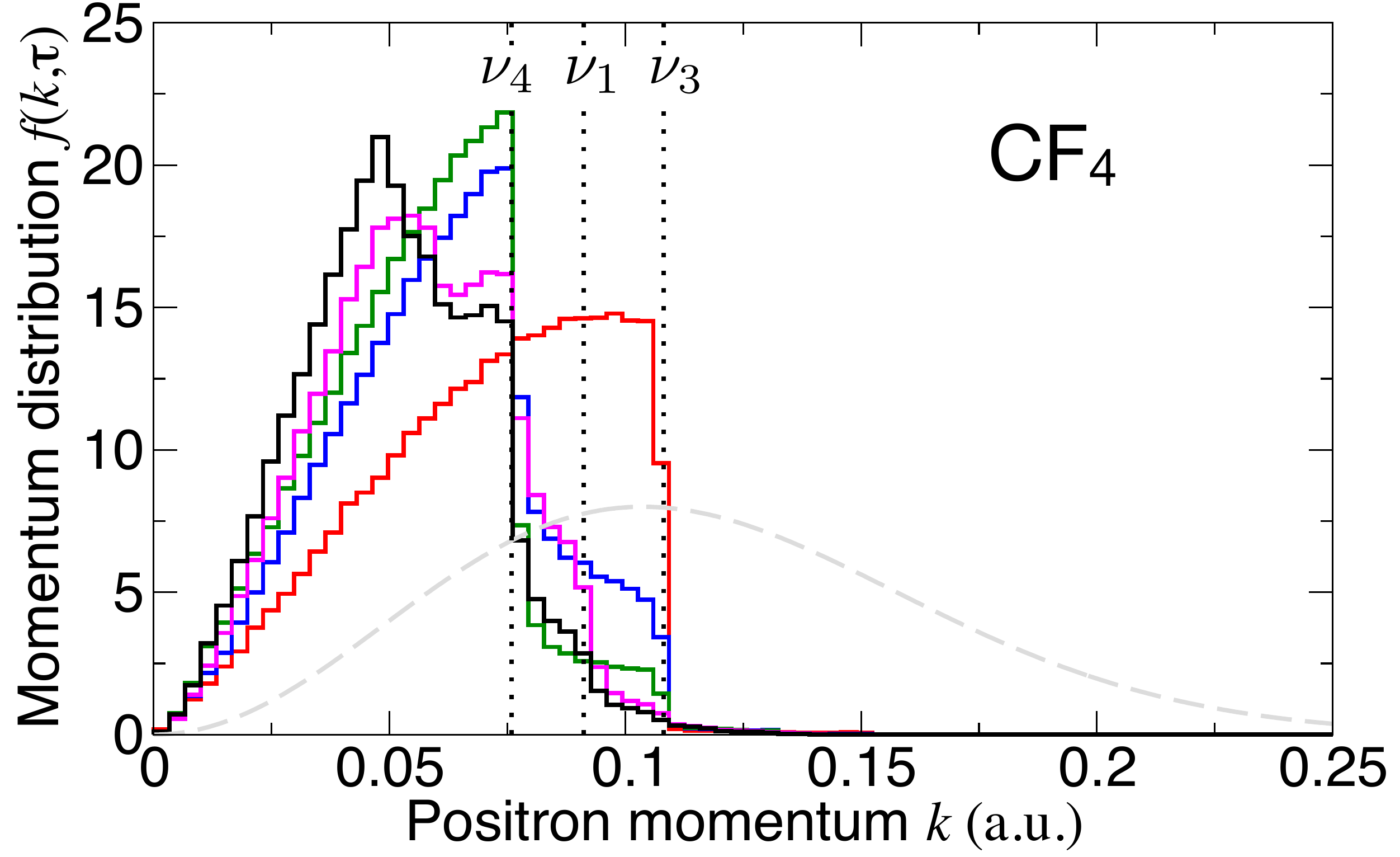}
%\caption{\label{fig:CF4_momdist_approximations}Positron momentum distribution $f(k,\tau)$ at $\tau=0.06$~ns~amg for cooling in CF$_4$. 
%Solid lines: $\nu_3$, $\nu_3+\nu_4$, $\nu_3+4\nu_4$, $\nu_3+\nu_4+\nu_1$,  $\nu_3+4\nu_4+4\nu_1$ approximations (see text; colored as in Fig.~\ref{fig:CF4_temperature}).
%Dashed gray line: MB for $T=1700$~K. 
%Dotted lines:  vibrational excitation thresholds.
\caption{\label{fig:CF4_momdist_approximations}Calculated PMD $f(k,\tau)$ at $\tau=0.06$~ns~amg for cooling in CF$_4$ in the $\nu_3$, $\nu_3+\nu_4$, $\nu_3+4\nu_4$, $\nu_3+\nu_4+\nu_1$,  and $\nu_3+4\nu_4+4\nu_1$ approximations (see text; colored as in Fig.~\ref{fig:CF4_temperature}).
Also shown is the initial MB for $T=1700$~K (dashed-gray line) and the vibrational excitation thresholds (dotted vertical lines). 
}
\end{figure}

Figure \ref{fig:CF4_temperature} shows the calculated time-dependent positron temperature %, cooling from an initial MB at 1700 K, 
compared with experiment  \cite{Natisin14,CF4_exp_note}. 
The simulation used 200\,000 positrons with $\Delta\tau=2\times10^{-6}$~ns~amg.
The figure separately shows the results for when only the $0\rightleftarrows3$ transitions are included; when the $0\rightleftarrows3$ and $0\rightleftarrows4$ transitions are included; and when the $0\rightleftarrows3$, $0\rightleftarrows4$, and $0\rightleftarrows1$ transitions are included (denoted  $\nu_3$; $\nu_3+\nu_4$; and $\nu_3+\nu_4+\nu_1$, respectively). 
Also shown are the results for when  $\sigma_{0\rightleftarrows4}$ and $\sigma_{0\rightleftarrows1}$  are scaled by the factor of 4 to approximate the coupled-channel calculations (denoted  $\nu_3+4\nu_4$ and $\nu_3+4\nu_4+4\nu_1$, respectively).
The $\nu_3+4\nu_4+4\nu_1$ calculation gives excellent agreement with experiment.
The other approximations predict slower cooling of the positrons.
In fact, for $\nu_3$ and $\nu_3+\nu_4$, the positrons do not thermalize close to $T_{ R}=300$~K. 
The reason for this is most easily seen by considering $f(k,\tau)$ for a particular value of $\tau$. %  in the various approximations. 
Figure \ref{fig:CF4_momdist_approximations} shows $f(k,\tau)$ in the various approximations for $\tau=0.06$~ns~amg [see also the video {\tt \href{https://youtu.be/BHJ5ezj1_lE}{CF4-video}} in Supplemental Material \cite{CF4_video}].
Also shown is the MB for $T=1700$~K from which the positron momenta at $\tau=0$ are sampled (dashed gray line), the peak of which almost coincides with the value of $k$ corresponding to the $0\to3$ excitation threshold energy (rightmost dotted black line).
In the $\nu_3$ approximation, only  positrons above %whose initial energies are greater than 
this threshold can lose energy. Thus, at $\tau=0.06$~ns~amg, when equilibrium has been reached, we observe a pileup in $f(k,\tau)$ just below this threshold, with $f(k,\tau)\sim 0$ above the threshold. 
The positron temperature can thus decrease no further.
In the $\nu_3+\nu_4$ and $\nu_3+4\nu_4$ approximations, at $\tau=0.06$~ns~amg, we observe pileups below both thresholds. 
Both diminish on longer timescales than considered here: those just below the $0\to3$ threshold can cool via $0\to4$ excitation; those below the $0\to4$ threshold can cool, but  only via \emph{multiple}  $4\to0$ deexcitations (since $2\varepsilon_4<\varepsilon_3$) followed by a $0\to3$ excitation. 
The $\nu_3+\nu_4+\nu_1$ and $\nu_3+4\nu_4+4\nu_1$ approximations are notably different. %Importantly, 
The additional $\nu_1$ mode %facilitates more efficient cooling,
 enables positrons %whose energies are smaller than 
 below the lowest ($0\to4$) excitation threshold to cool further via a \emph{single} deexcitation and excitation of the molecule. 
For example, a positron with energy $\varepsilon$ where $\varepsilon_1-\varepsilon_4<\varepsilon<\varepsilon_4$ can induce a $4\to0$ deexcitation followed by a $0\to1$ excitation, thus reducing its energy by $\varepsilon_1-\varepsilon_4$; another pathway is via a $1\to0$ deexcitation followed by a $0\to3$ excitation. 
Since the deexcitation cross sections are orders of magnitude smaller than the excitation ones above the vibrational excitation thresholds \cite{supp_info}, such cooling via a single deexcitation and excitation is much more probable than via multiple deexcitations followed by excitation. 
Indeed, the peak in $f(k,\tau)$ (at $k\approx 0.05$ a.u.) corresponds to the energy $\varepsilon_1-\varepsilon_4$, below which positrons can cool further only via the improbable multiple deexcitations pathway. 
%Overall, 
The doorway provided by the $\nu_1$ mode thus appears to provide an accurate cooling timescale and room-temperature thermalization.

\begin{figure}
\centering
\includegraphics*[width=0.875\columnwidth]{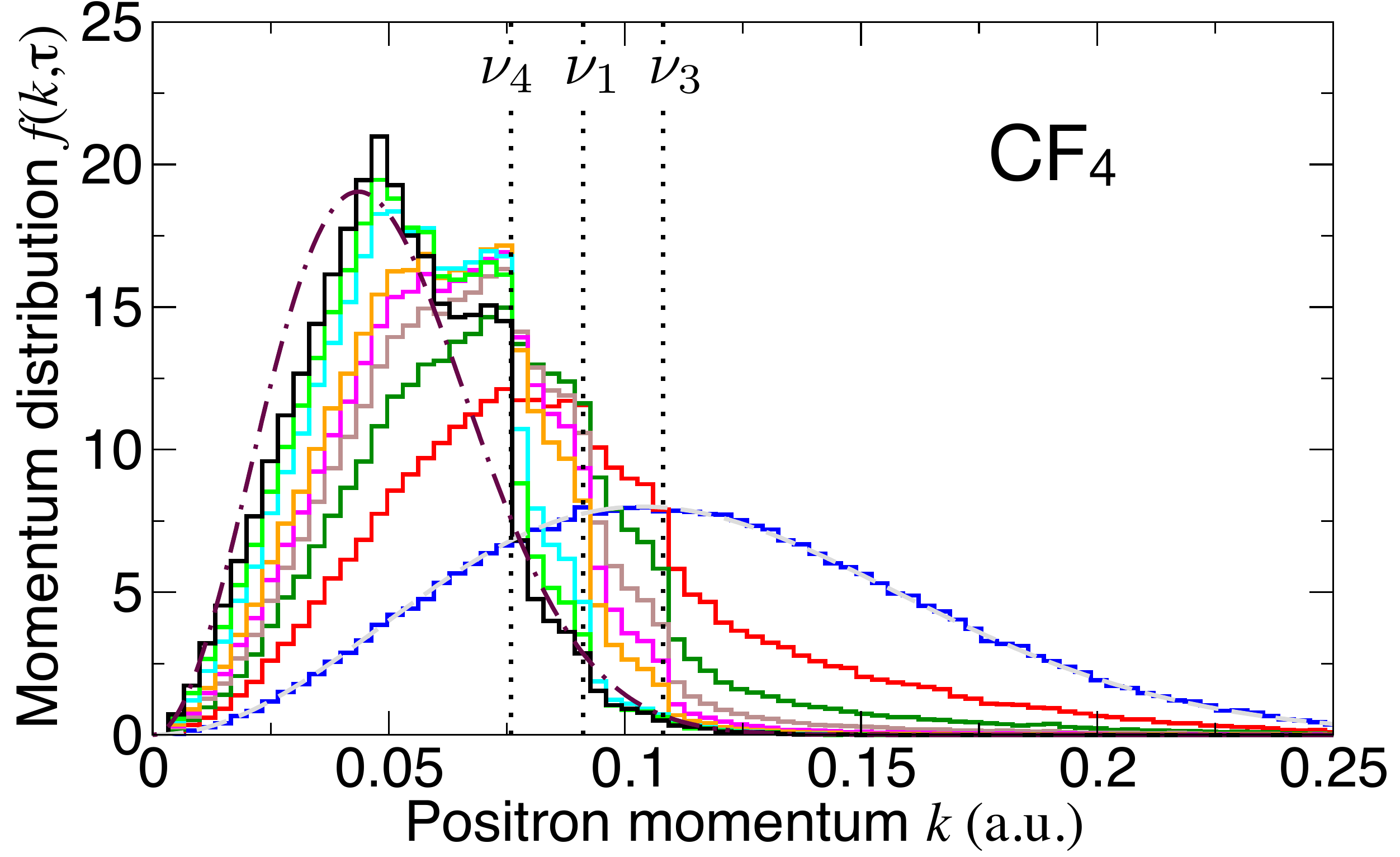}
%\caption{\label{fig:CF4_momdist_best}Positron momentum distribution  $f(k,\tau)$   for cooling in CF$_4$, in the $\nu_3+4\nu_4+4\nu_1$  approximation. 
%Solid lines: $f(k,\tau)$ for values of $\tau$ as follows (in ns~amg): 
%blue, 0; red, 0.002; dark green, 0.004; brown, 0.006; magenta, 0.008; orange, 0.01; cyan, 0.02; light green, 0.03; black, 0.06.
%Dashed gray line: MB for $T=1700$~K. 
%Dot-dashed line: MB for $T_{\rm R}=300$~K. 
%Dotted vertical lines: vibrational excitation thresholds. 
%}
\caption{\label{fig:CF4_momdist_best}Calculated PMD  $f(k,\tau)$   for cooling in CF$_4$, in the $\nu_3+4\nu_4+4\nu_1$ approximation at the following values of $\tau$ (in ns~amg):
%Solid lines: $f(k,\tau)$ for values of $\tau$ as follows (in ns~amg): 
0 (blue),
0.002 (red), 
0.004 (dark green), 
0.006 (brown), 
0.008 (magenta), 
0.01 (orange),
0.02 (cyan), 
0.03 (light green), 
0.06 (black).
Also shown is the initial MB for $T=1700$~K (dashed gray line), the MB for $T_{\rm R}=300$~K (dot-dashed line) and the vibrational excitation thresholds (dotted vertical lines). 
}
\end{figure}

\begin{figure}
\centering
\includegraphics*[width=0.875\columnwidth]{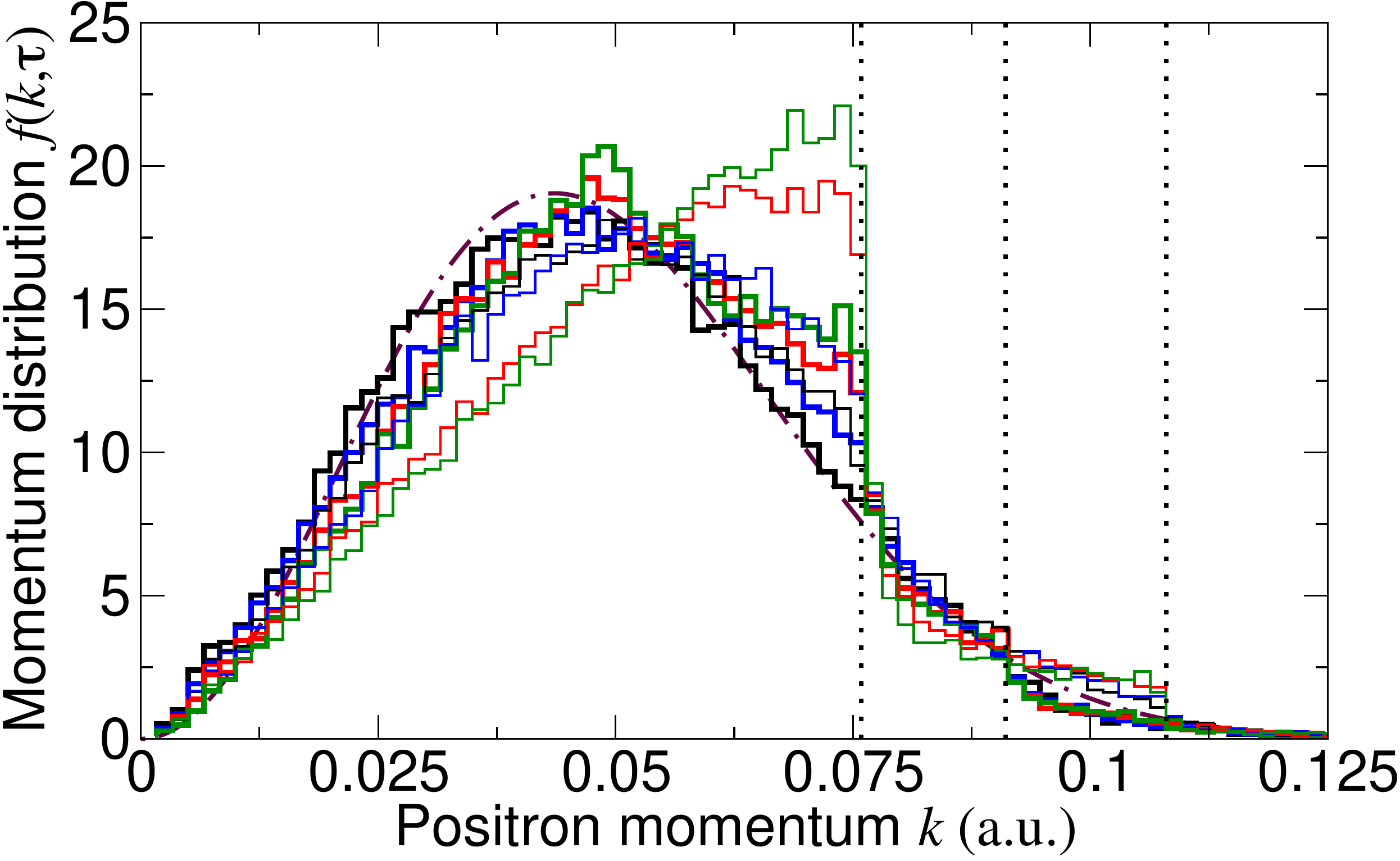}
\caption{\label{fig:CF4_momdist_approximations_pospos}Calculated PMD $f(k,\tau)$ at $\tau=0.06$~ns~amg for cooling in CF$_4$ with the inclusion of positron-positron interactions. Thin lines, $\nu_3+4\nu_4$ approximation; thick lines, $\nu_3+4\nu_4+4\nu_1$ approximation. Values of $R$: $1\times10^{-6}$ (black), $5\times10^{-7}$ (blue), $1\times10^{-7}$ (red), $1\times10^{-8}$ (green). Dot-dashed line, MB for $T_R=300$~K.}
\end{figure}

\begin{figure}
\centering
\includegraphics*[width=0.875\columnwidth]{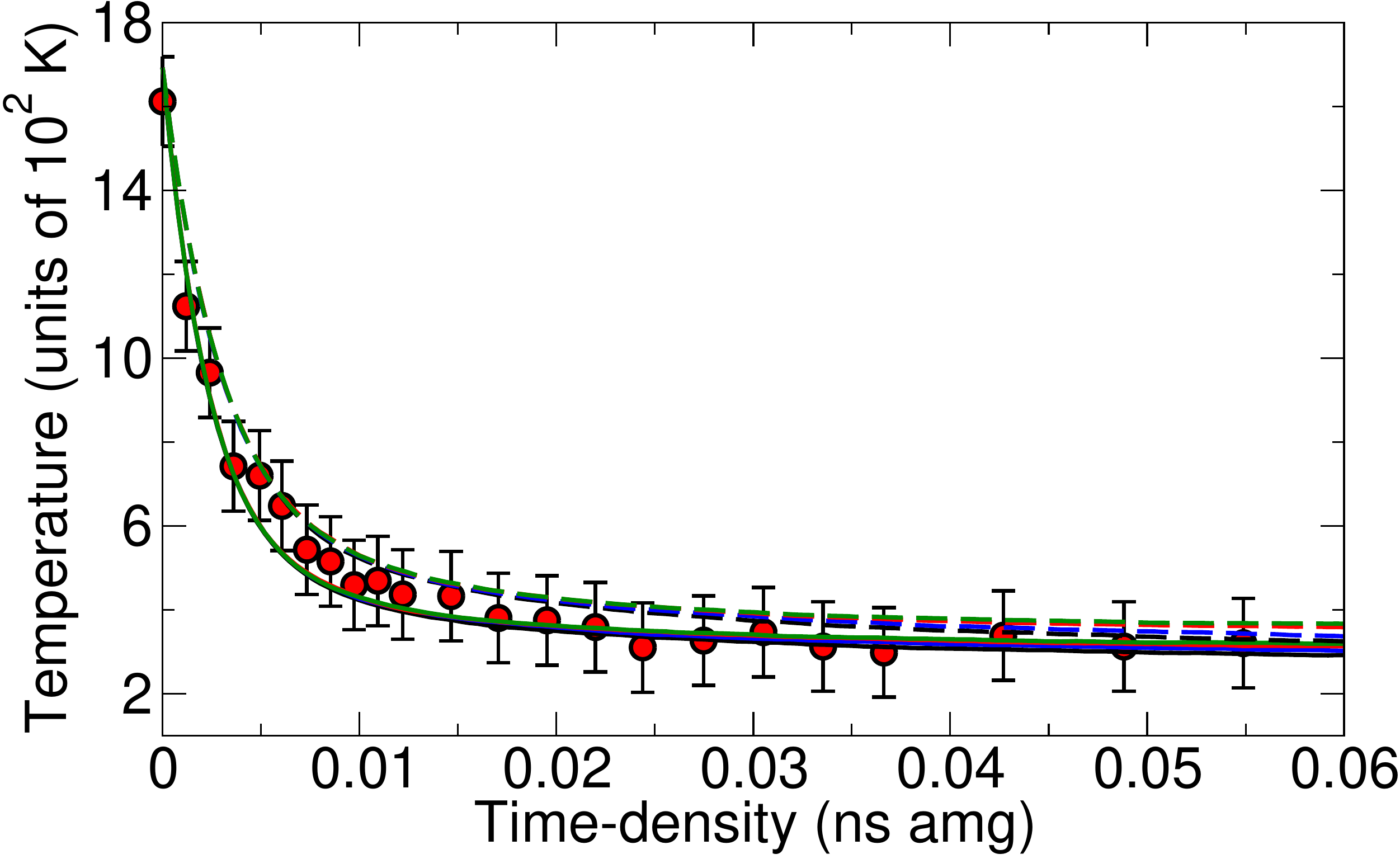}
\caption{\label{fig:CF4_temp_pospos}Positron temperature during cooling in CF$_4$ with the inclusion of positron-positron interactions. Dashed lines, $\nu_3+4\nu_4$ approximation; solid lines, $\nu_3+4\nu_4+4\nu_1$ approximation (colored as in Fig.~\ref{fig:CF4_momdist_approximations_pospos}). Red circles, experiment \cite{Natisin14}.}
\end{figure}

%\red{In all of the approximations, $f(k,\tau)$ notably deviates  from the initial MB. This is because in inducing vibrational transitions of the molecule, the positrons lose or gain energy in large amounts relative to the overall energy spread of the distribution, e.g., the energy lost or gained in  a $0\rightleftarrows4$ transition is ${\sim}10$\% of the initial overall spread of positron energies (based on an assumed initial momentum spread of $0\leq k\leq 0.25$~a.u.).}
In all of the approximations, the PMD at equilibrium is markedly non-Maxwellian. This is because in inducing the vibrational transitions, the positrons lose or gain energy in large amounts relative to the overall energy spread of the PMD, e.g., the energy lost  in  a $0\to4$ excitation is ${\sim}10$\% of the initial overall spread of positron energies (assuming an initial momentum spread of $0\leq k\leq 0.25$~a.u.; see Fig.~\ref{fig:CF4_momdist_approximations}).
Figure \ref{fig:CF4_momdist_best} shows 
 $f(k,\tau)$ for the best approximation, $\nu_3+4\nu_4+4\nu_1$, for several $\tau$ [see also the video {\tt \href{https://youtu.be/BHJ5ezj1_lE}{CF4-video}} in Supplemental Material \cite{CF4_video}].
The PMD for $\tau=0.06$~ns~amg (the final value of $\tau$ displayed), though  close in shape and magnitude to the MB for $T_{ R}=300$~K, has a lingering pileup at the $0\to4$ threshold, precluding full Maxwellianization of the PMD. 
%which would continue to diminish for greater $\tau$.
%Since cooling with positron-molecule interactions alone causes $f(k,\tau)$ to be non-Maxwellian, 
We now consider, therefore, the possible role of positron-positron collisions. Figure \ref{fig:CF4_momdist_approximations_pospos} shows $f(k,\tau)$ for $\tau=0.06$~ns~amg in the $\nu_3+4\nu_4$ and $\nu_3+4\nu_4+4\nu_1$ approximations, using 50\,000 positrons, with the inclusion of positron-positron interactions, for  $R=10^{-8}$--$10^{-6}$. 
For $R=10^{-6}$,  the positron-positron collisions clearly  Maxwellianize  the PMD in both approximations, eliminating the pileup at the $0\to4$ threshold almost completely. 
For $R=10^{-7}$, however, the effect of positron-positron collisions is much smaller, and
for $R=10^{-8}$, the PMD in both approximations is essentially the same as it was without positron-positron collisions (cf.~Fig.~\ref{fig:CF4_momdist_approximations}).
Thus, in the experiment of Natisin \textit{et al.} \cite{Natisin14}, since $R$ is estimated to be ${\sim}10^{-7}$--$10^{-6}$  \cite{Surko_private}, positron-positron collisions may or may not play a significant role.
Figure \ref{fig:CF4_temp_pospos} shows the time dependence of the  positron temperature in the $\nu_3+4\nu_4$ and $\nu_3+4\nu_4+4\nu_1$ approximations, with the inclusion of positron-positron interactions.
For all values of $R$ considered, even  $R=10^{-6}$, the $\nu_3+4\nu_4+4\nu_1$ approximation  provides  slightly better agreement with experiment than the $\nu_3+4\nu_4$ approximation.

% \red{[(Assuming new results dont change much from the old ones) we need to make clear that on their own the pos-pos interactions are insufficient, and don't provide thermalisation to room temp nor give a cooling curve agreement with experiment... i.e., its a sweet spot where both $nu_1$ and pos-pos are effective]}

\emph{Positron cooling in N$_2$ gas}.---The homonuclear diatomic N$_2$ molecule has no permanent electric dipole moment, and its fundamental vibrational mode is  dipole inactive.  
Therefore, we initially assume that positron cooling in N$_2$ proceeds via quadrupole rotational excitations. 
To a good approximation, the energy of a rotational level with angular momentum $J$ is $\varepsilon_J=BJ(J+1)$, where $B$ is the rotational constant. For N$_2$, we take $B=9.2\times10^{-6}$~a.u. \cite{Huber79}.
We use the Born-quadrupole  cross sections for the $J\rightleftarrows J+2$ transitions %,  
%which are are given by the Gerjuoy-Stein equations 
\cite{Gerjuoy55},
%\begin{align}
%\sigma_{J\to J+2}(k) &= \frac{8\pi}{15} Q^2 \frac{(J+1)(J+2)}{(2J+1)(2J+3)} \sqrt{1 - \frac{2\,\Delta \varepsilon_J}{k^2}} , \label{eq:GS_ex} \\
%\sigma_{J+2\to J}(k) &= \frac{8\pi}{15} Q^2 \frac{(J+1)(J+2)}{(2J+3)(2J+5)} \sqrt{1 + \frac{2\,\Delta \varepsilon_J}{k^2}} , \label{eq:GS_deex}
%\end{align}
%where $Q$ is the molecular quadrupole moment, which we take as $Q=1.27$~a.u. \cite{Graham98}, and $\Delta \varepsilon_J = \varepsilon_{J+2}-\varepsilon_J= 2B(2J+3)$ 
%\footnote{Also, the degeneracy of a rotational level, required to determine its Boltzmann-averaged occupation number in equilibrium, is $g_J=6(2J+1)$ for even $J$, or $g_J=3(2J+1)$ for odd $J$.}.
%For room temperature N$_2$, the Boltzmann occupation number falls off rapidly above $J=20$ (that for $J=30$ is $\sim$2 orders of magnitude lower than the $J=20$), 
and we find that including up to $J=60$ gives converged results \cite{supp_info}.
We neglect the role of positron-positron collisions.

\begin{figure}
\centering
\includegraphics*[width=0.885\columnwidth]{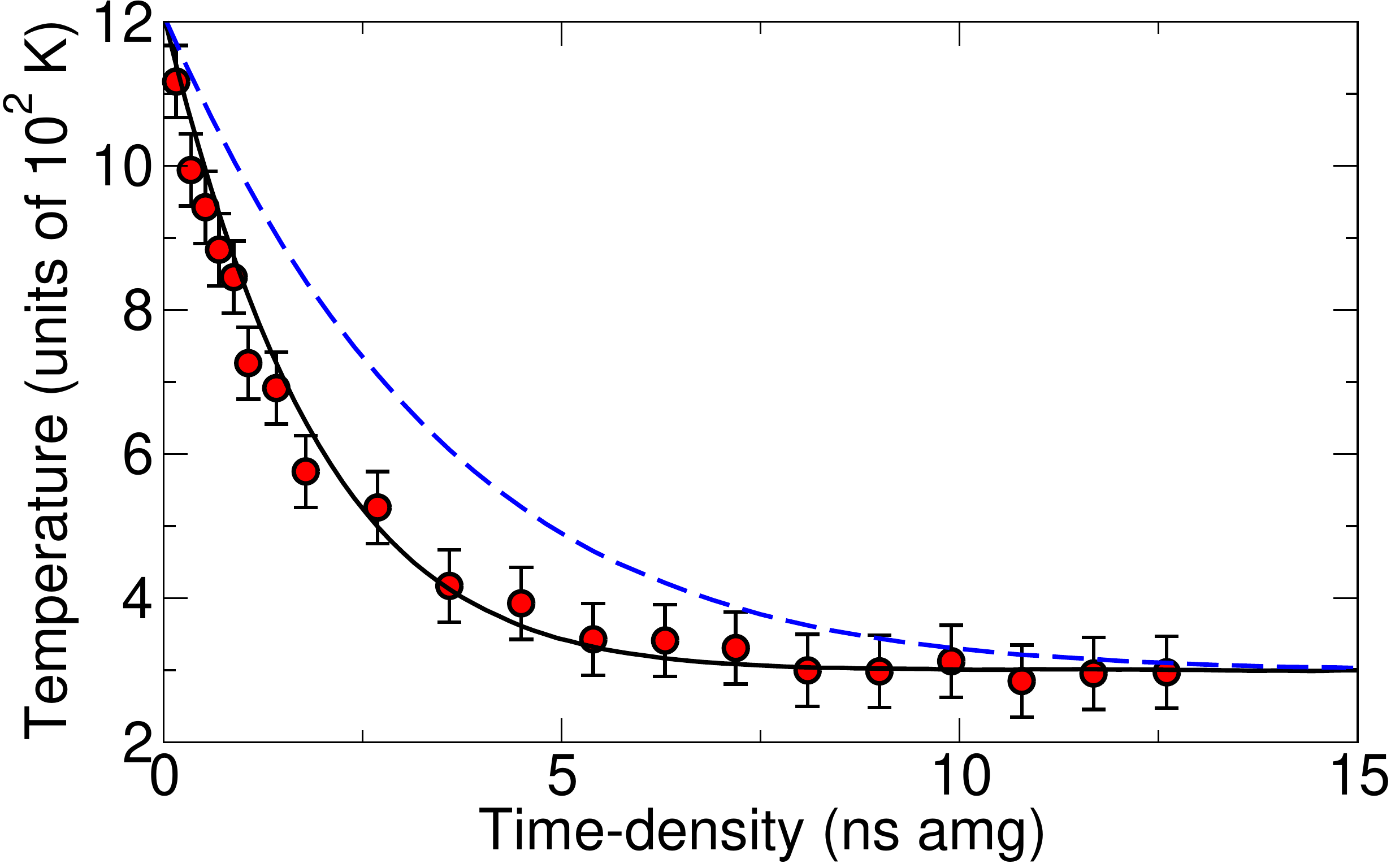}
%\caption{\label{fig:N2_temp}Positron temperature during cooling in N$_2$. 
%Lines: calculations for  values of $J_\text{max}$ as follows: dashed magenta, 10; dotted red, 15; dot-dashed green, 20; solid black, 60; dot-dash-dotted blue, 60 (with unscaled cross sections).
%Red circles: experimental data \cite{Natisin14}.}
%\caption{\label{fig:N2_temp}Positron temperature during cooling in N$_2$. 
%Calculations for $J_\text{max} = $ 10 (dashed magenta), 15 (dotted red), 20 (dot-dashed green), 60 (solid black), and 60 with unscaled cross sections (dot-dash-dotted blue), compared with experiment \cite{Natisin14} (red circles).}  
\caption{\label{fig:N2_temp}Positron temperature during cooling in N$_2$. 
Calculations with $J_\text{max}=60$ using  unscaled cross sections  (dashed blue) and  scaled cross sections   (solid black); experiment \cite{Natisin14} (red circles).}  
\end{figure}

Figure \ref{fig:N2_temp} shows the calculated time-dependent positron temperature during cooling from an initial MB at 1200~K (corresponding to the experiment \cite{Natisin14,N2_exp_note}). 
We use 200\,000 positrons and $\Delta\tau=5\times10^{-4}$~ns~amg.
The calculation (dashed blue line) %using Eqs.~(\ref{eq:GS_ex}) and (\ref{eq:GS_deex})
 predicts  slower cooling than the experiment (red circles). A similar phenomenon was observed in the theoretical model of Natisin \textit{et al.} \cite{Natisin14}, suggesting that the Born-quadrupole approximation may underestimate the true rotational excitation cross sections. Natisin \textit{et al.} found that scaling the Born-quadrupole cross sections by an empirical factor of 1.8 gave much better agreement between the predictions of their model and their experimental data \cite{Natisin14}, and yielded a magnitude for $\sigma_{0\to2}$ in better agreement with calculation \cite{PhysRevA.43.2538}.
Thus, we likewise performed calculations with the Born-quadrupole cross sections scaled by %this factor of
 1.8 (solid black line in Fig.~\ref{fig:N2_temp}).
%Figure~\ref{fig:N2_temp} also shows $T(\tau)$ as obtained using the scaled cross sections. 
Excellent agreement with the experiment is achieved; the thermalization time is consistent with the 14 ns amg measured in Ref.~\cite{AlQaradawi:2000}.

\begin{figure}
\centering
\includegraphics*[width=0.885\columnwidth]{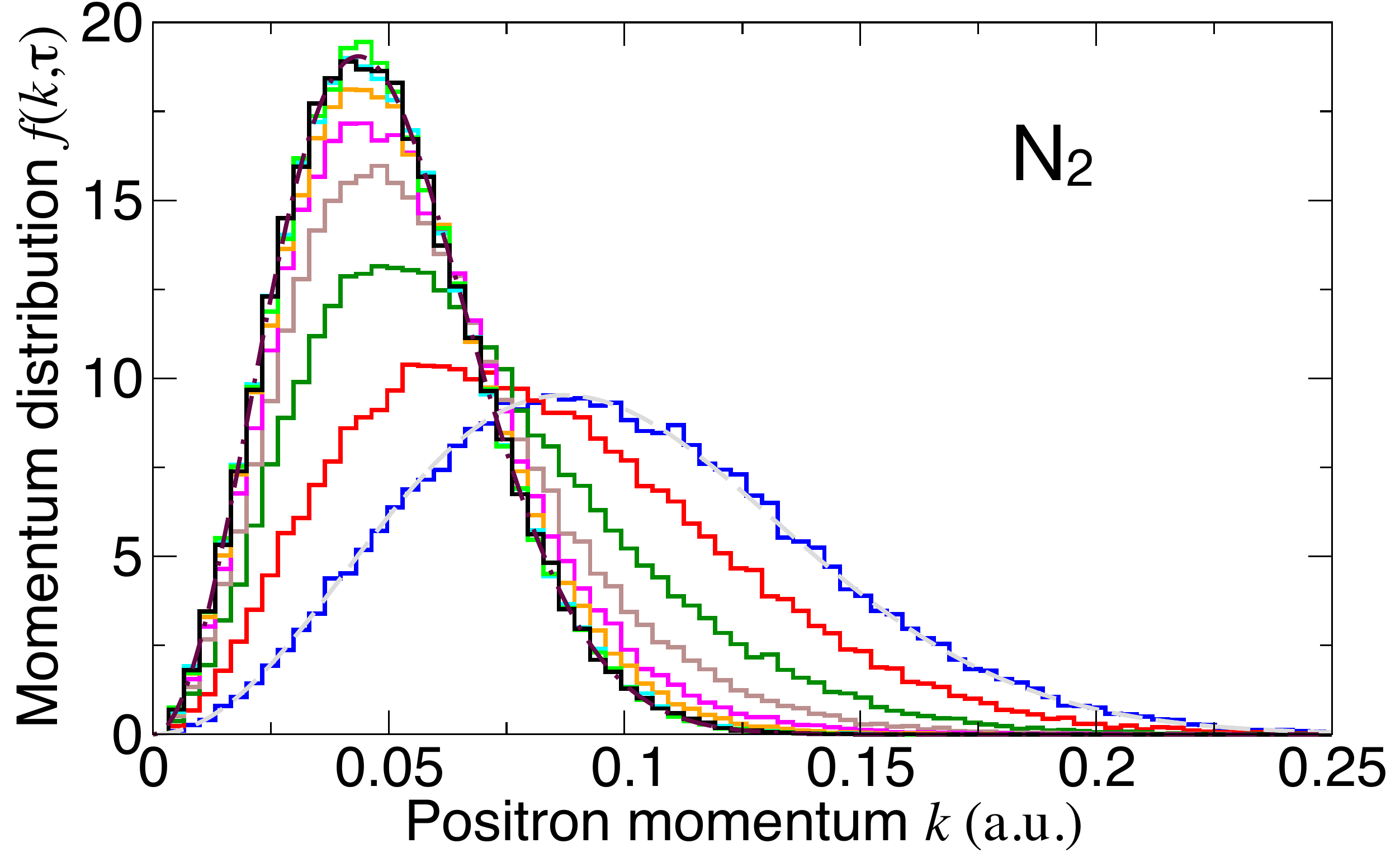}
%\caption{\label{fig:N2_momdist}Positron momentum distribution  $f(k,\tau)$   for cooling in N$_2$.
%Solid lines: $f(k,\tau)$ for  values of $\tau$ as follows (in ns~amg): blue, 0; red, 0.9; dark green, 2.1; brown, 3.3; magenta, 4.5; orange, 5.7; cyan, 8.1; light green, 10.5; black, 12.9.
%Dashed gray line: MB for $T=1200$~K. 
%Dot-dashed maroon line: MB for $T_{\rm R}=300$~K.
\caption{\label{fig:N2_momdist}Positron momentum distribution  $f(k,\tau)$   for cooling in N$_2$ for $\tau$ as follows (in ns~amg): 
0 (blue),
0.9 (red), 
2.1 (dark green),
3.3 (brown),
4.5 (magenta), 
5.7 (orange),
8.1 (cyan),
10.5 (light green),
12.9 (black). 
Also shown in the initial MB for $T=1200$~K (dashed gray line) and the MB for $T_{\rm R}=300$~K (dot-dashed maroon line).
}
\end{figure}

Figure \ref{fig:N2_momdist} shows $f(k,\tau)$ for several values of  $\tau$ [see also the video {\tt \href{https://youtu.be/wR1zdg_9gtc}{N2-video}} in Supplemental Material \cite{N2_video}].
In contrast to CF$_4$, $f(k,\tau)$ for N$_2$ remains  near-Maxwellian as $\tau$ proceeds. Moreover,  $R$ is ${\sim}30$ times smaller in the experiment for N$_2$ than in the experiment for CF$_4$, so  positron-positron collisions will not play a significant role. 
Indeed,  for the final value of $\tau$ shown, viz., $\tau=12.9$~ns~amg, $f(k,\tau)$ follows a MB for $T_{ R}=300$~K very closely. This is a result of the relatively small energy spacing between the various rotational levels in comparison to that between the vibrational levels in CF$_4$.

%\red{Finally, we briefly investigated the effect of accounting for possible excitation of the dipole-inactive vibrational mode \red{State its energy to make more illuminating}. We estimated the relevant cross sections using existing  close-coupling electron-impact calculations  \cite{Robertson97}  but found  that the cooling was not significantly affected \cite{supp_info}.}

%\blue{positron-positron: It is known that positron-positron interactions can rethermalize the distribution much faster than the vibrational cooling timescale. 
%\cite{RevModPhys.87.247,Trunec03}}.
 
\textit{Conclusions}.---Positron cooling via inelastic collisions in CF$_4$ and N$_2$ gases---the buffer gases of choice in the ubiquitous Surko traps---has been studied via a Monte Carlo approach. Specifically, we simulated the experiment of Natisin \emph{et al.} \cite{Natisin14}, where positrons cooled following microwave heating to ${\sim}1500$~K.  Cooling in CF$_4$ (N$_2$) was assumed to proceed via vibrational (rotational) excitations. 
%\red{Born-approximation transition cross sections (with some empirical scaling factors, which for CF$_4$ enabled approximation of coupled-channel cross sections) were used. }
For CF$_4$, we found that including the two dipole-active modes $\nu_3$ and $\nu_4$ was insufficient. 
Because of the relatively large amounts of energy lost by the positron in inducing a discrete vibrational excitation, the PMD does not remain Maxwellian: pileups  are observed near each of the vibrational excitation thresholds. We found that inclusion of the $\nu_1$ mode can provide a doorway for further cooling, diminishing the pileups below the dipole-active thresholds and ultimately providing excellent agreement with experiment for the time dependence of the positron temperature. However, even with the inclusion of $\nu_1$, the PMD remains markedly non-Maxwellian. We found that positron-positron collisions can effect efficient Maxwellianization of the PMD for $R=10^{-6}$, but not for $R=10^{-7}$ (or smaller), where $R$ is the ratio of the positron and gas number densities. In the experiment, $R\sim10^{-7}$--$10^{-6}$ is estimated \cite{Surko_private}, so the importance of positron-positron collisions in the experiment is unclear.
For N$_2$, the energy spacing between the rotational levels is much smaller than that between the vibrational levels of CF$_4$, so the distribution remains near-Maxwellian throughout the cooling.
%\red{For both gases, the excellent agreement with experiment provides verification of the cross sections, including the coupled-channel cross sections for CF$_4$ \cite{Franz09}. }
Improved calculations or measurements of the positron-impact transition cross sections between rovibrational levels would enable more definitive calculations of cooling. In principle, annihilation of positrons during the cooling is another process that can affect the overall PMD and positron temperature, but this is not expected to play a significant role for cooling in CF$_4$ and N$_2$
%In particular, a combined many-body theory and coupled-channel approach may enable \emph{ab initio} calculations of the inelastic and annihilation cross sections 
\footnote{The annihilation cross section is $\sigma_\text{ann}=\pi r_0^2 c Z_\text{eff}/v$, where $r_0$ is the classical electron radius, $c$ is the speed of light, $v$ is the positron velocity, and $Z_\text{eff}$ is the effective number of electrons contributing to annihilation. Measurements and calculations of $Z_\text{eff}$ for CF$_4$ and N$_2$  indicate $Z_\text{eff}\sim10^1$ for both species  \cite{Charlton85,Marler04,Swann20}, which makes $\sigma_\text{ann}$ negligible in comparison to the total vibrational (for CF$_4$) or rotational (for N$_2$) cross section \cite{supp_info}.}.

%\emph{Data availability}.---All relevant data generated and analyzed during this work, and the modified {\sc anticool} computer program used to produce it, are presently available from the authors on reasonable request. On acceptance, the data will be made freely available on the Queen’s University Belfast official data repository https://pure.qub.ac.uk/en/datasets/.  

\begin{acknowledgments}
\emph{Acknowledgments}.---We thank Cliff Surko and James Danielson for useful discussions regarding the experiment, and Gleb Gribakin for encouraging us to consider positron-positron interactions and providing useful comments on the draft manuscript. 
This work was supported by ERC StG 804383 ``ANTI-ATOM''. %D.G.G.'s European Research Council StG 804383 ``ANTI-ATOM''.
\end{acknowledgments}

\vspace*{-4ex}
%\bibliography{pos_mol_cool_bib}

\begin{thebibliography}{49}
\bibitem [{CF4({\natexlab{a}})}]{CF4_existing_note}%
  \BibitemOpen
  \href@noop {} {\bibinfo {title} {As far as we are aware, the only existing
  calculations of positron-impact vibrational excitation of {CF$_4$} are the
  {B}orn-dipole calculation \cite{Itikawa74} and a semiempirical
  coupled-channel calculation \cite{Franz09}. for positron-impact rotational
  excitation of {N$_2$}, apart from the {B}orn-quadrupole calculation
  \cite{Gerjuoy55}, other calculations present only cross sections for
  excitations {$0\to J$} for a few values of the angular momentum {$J$} or
  summed cross sections \cite{Mukherjee15,Barp18a}.}}
  ({\natexlab{a}})\BibitemShut {NoStop}%
\end{thebibliography}
\begin{thebibliography}{49}
\bibitem [{CF4({\natexlab{a}})}]{CF4_existing_note}%
  \BibitemOpen
  \href@noop {} {\bibinfo {title} {As far as we are aware, the only existing
  calculations of positron-impact vibrational excitation of {CF$_4$} are the
  {B}orn-dipole calculation \cite{Itikawa74} and a semiempirical
  coupled-channel calculation \cite{Franz09}. for positron-impact rotational
  excitation of {N$_2$}, apart from the {B}orn-quadrupole calculation
  \cite{Gerjuoy55}, other calculations present only cross sections for
  excitations {$0\to J$} for a few values of the angular momentum {$J$} or
  summed cross sections \cite{Mukherjee15,Barp18a}}}
\BibitemShut {NoStop}%
\end{thebibliography}

\end{document}